\documentclass[twocolumn]{aastex63} 

\usepackage{amsmath}
\usepackage{xspace}
\usepackage{booktabs}

\usepackage{floatrow}
\usepackage{natbib}

\DeclareFloatVCode{somespace}{\vspace{1.667\baselineskip}}
\newcommand{\swift}{{\it Swift}\xspace}
\newcommand{\nicer}{\textit{NICER}\xspace}
\newcommand{\xmm}{{\it XMM}\xspace}
\newcommand{\eros}{{eROSITA}\xspace}
\newcommand{\target}{{AT2018fyk}\xspace}

\newcommand{\alphaox}{$\alpha_{\rm ox}$\,}

\shorttitle{AT2018FYK as a repeating partial stellar tidal disruption event}
\shortauthors{Wevers et al.}

\begin{document}

\title{Live to die another day: the rebrightening of AT2018fyk as a repeating partial tidal disruption event}
\correspondingauthor{Thomas Wevers}
\email{twevers@eso.org}
\author[0000-0002-4043-9400]{T. Wevers}
\affiliation{European Southern Observatory, Alonso de Córdova 3107, Vitacura, Santiago, Chile}
\author[0000-0003-3765-6401]{E.~R.~Coughlin}
\affiliation{Department of Physics, Syracuse University, Syracuse, NY 13210, USA}
\author[0000-0003-1386-7861]{D.R. Pasham}
\affiliation{Kavli Institute for Astrophysics and Space Research, Massachusetts Institute of Technology, Cambridge, MA, USA}
\author[0000-0002-5063-0751]{M. Guolo}
\affiliation{Department of Physics and Astronomy, Johns Hopkins University, 3400 N. Charles St., Baltimore MD 21218, USA}
\author{Y. Sun}
\affiliation{University of Arizona, 933 N. Cherry Ave., Tucson, AZ 85721, USA}
\author[0000-0002-0934-2686]{S. Wen}
\affiliation{Department of Astrophysics/IMAPP, Radboud University, P.O. Box 9010, 6500 GL, Nijmegen, The Netherlands}
\author[0000-0001-5679-0695]{P.G. Jonker}
\affiliation{Department of Astrophysics/IMAPP, Radboud University, P.O. Box 9010, 6500 GL, Nijmegen, The Netherlands}
\affiliation{SRON, Netherlands Institute for Space Research, Niels Bohrweg 4, 2333 CA, Leiden, The Netherlands}
\author[0000-0001-6047-8469]{A. Zabludoff}
\affiliation{University of Arizona, 933 N. Cherry Ave., Tucson, AZ 85721, USA}
\author[0000-0002-8851-4019]{A. Malyali}
\affiliation{Max-Planck-Institut f{\"u}r extraterrestrische Physik, Giessenbachstra{\ss}e, 85748 Garching, Germany}
\author[0000-0003-4054-7978]{R. Arcodia}
\affiliation{Max-Planck-Institut f{\"u}r extraterrestrische Physik, Giessenbachstra{\ss}e, 85748 Garching, Germany}
\author{Z. Liu}
\affiliation{Max-Planck-Institut f{\"u}r extraterrestrische Physik, Giessenbachstra{\ss}e, 85748 Garching, Germany}
\author[0000-0002-0761-0130]{A. Merloni}
\affiliation{Max-Planck-Institut f{\"u}r extraterrestrische Physik, Giessenbachstra{\ss}e, 85748 Garching, Germany}
\author[0000-0001-5990-6243]{A. Rau}
\affiliation{Max-Planck-Institut f{\"u}r extraterrestrische Physik, Giessenbachstra{\ss}e, 85748 Garching, Germany}
\author{I. Grotova}
\affiliation{Max-Planck-Institut f{\"u}r extraterrestrische Physik, Giessenbachstra{\ss}e, 85748 Garching, Germany}
\author[0000-0002-5096-9464]{P. Short}
\affiliation{Institute for Astronomy, University of Edinburgh, Royal Observatory, Blackford Hill, Edinburgh EH9 3HJ, UK}
\author{Z. Cao}
\affiliation{Department of Astrophysics/IMAPP, Radboud University, P.O. Box 9010, 6500 GL, Nijmegen, The Netherlands}
\begin{abstract}
Stars that interact with supermassive black holes (SMBHs) can either be completely or partially destroyed by tides. In a partial tidal disruption event (TDE) the high-density core of the star remains intact, and the low-density, outer envelope of the star is stripped and feeds a luminous accretion episode. 
The TDE AT2018fyk, with an inferred black hole mass of $10^{7.7\pm0.4}$ M$_{\odot}$, experienced an extreme dimming event at X-ray (factor of $>$6000) and UV (factor $\sim$15) wavelengths $\sim$500--600 days after discovery. Here we report on the re-emergence of these emission components roughly 1200 days after discovery. 
We find that the source properties are similar to those of the pre-dimming accretion state, suggesting that the accretion flow was rejuvenated to a similar state. We propose that a repeated partial TDE, where the partially disrupted star is on a $\sim 1200$ day orbit about the SMBH and is periodically stripped of mass during each pericenter passage, powers its unique lightcurve. This scenario provides a plausible explanation for AT2018fyk's overall properties, including the rapid dimming event and the rebrightening at late times. We also provide testable predictions for the behavior of the accretion flow in the future: if the second encounter was also a partial disruption then we predict another strong dimming event around day 1800 (August 2023), and a subsequent rebrightening around day 2400 (March 2025). This source provides strong evidence of the partial disruption of a star by a SMBH.
\end{abstract}


\section{Introduction}
\label{sec:introduction}
The classic prediction for the mass fall-back rate generated by a star being 
tidally disrupted by a supermassive black hole (SMBH) is an asymptotic, $t^{-5/3}$ decay \citep{Rees1988, Phinney1989}. While some of the tidal disruption events (TDEs) identified so far have displayed such long-term behavior, a significant fraction show different lightcurve evolution, which in some cases is completely decoupled from the mass fallback rate (e.g. \citealt{Gezari17, Kajava20}) and which may be expected (e.g. \citealt{Guillochon2013, Hayasaki21}). \citet{Auchettl2017} found that the X-ray lightcurves can be well described by power-law indices ranging from --0.5 to --2. \citet{Hammerstein22} defined three types of behavior for the UV/optical lightcurves, labeling them power-law decay (with indices ranging from --1 to --3, and a sizeable fraction that decay consistent with a t$^{-5/3}$ law), plateau and structured lightcurves. Deviations from the late-time t$^{-5/3}$ decay have also been suggested theoretically: \citet{Hayasaki13} and \citet{ Cufari2022a} found that stars on eccentric orbits can lead to a prompt shutoff in the lightcurve, while \citet{Guillochon2013} found that partial TDEs -- in which the dense stellar core survives the tidal encounter with the SMBH -- can lead to significant deviations from t$^{-5/3}$. \citet{Coughlin2019} predicted that partial TDEs should generically exhibit a $t^{-9/4}$ decay. More dramatic deviations, including truncation, order of magnitude dips and reflaring, can be induced by TDEs in SMBH binaries (e.g. \citealt{Liu09, Ricarte16, Coughlin17}). Recently, the source ASASSN-14ko was interpreted to be a \emph{repeating} partial TDE, such that the star is on a bound orbit about the SMBH and partially stripped of its mass -- thus feeding a new accretion flare -- each pericenter passage \citep{Payne2021}. 

In this work we report on the renewed X-ray and UV activity of the transient AT2018fyk, a proposed TDE originally described in \citet{Wevers19b}, $\sim$1200 days after discovery. We compare the observational properties to those of the previously observed accretion flow properties in Section \ref{sec:rebrightening}, after which we explore a repeating partial TDE scenario in Section \ref{sec:explanation} to explain the long-term properties. We present the implications and predictions of our model in Section \ref{sec:implications} before summarizing and concluding in Section \ref{sec:summary}.

\section{Long-term evolution of the X-ray and UV emission}
\label{sec:rebrightening}
\subsection{A brief history and basic properties of AT2018fyk}
ASASSN--18ul/AT2018fyk was discovered by the All-Sky Automated Survey for Supernovae \citep{Shappee14} on 2018 September 8 (MJD = 58369.2) in the nucleus of a galaxy (astrometric offset from the host galaxy center of light of 17$\pm$66 pc, \citealt{Wevers19b, Hodgkin21}) at a redshift of 0.059$\pm$0.0005. This corresponds to a luminosity distance of 274 Mpc by adopting a standard $\Lambda$CDM cosmology with H$_{0}$ = 67.4 km~s$^{-1}$~Mpc$^{-1}$, $\Omega_{m}$ = 0.315 and $\Omega_{\Lambda}$ = 1 - $\Omega_{m}$ = 0.685 \citep{planck}. Its classification as a TDE was based primarily on time-series of optical spectra, showing broad H, He as well as narrow Fe\,{\small II} and potentially N/O Bowen lines that evolved over time \citep{Wevers19b}; in addition, the host galaxy does not display any obvious narrow or other AGN-related emission lines (see section \ref{sec:hostagn} for further details). \citet{Wevers20} derived a SMBH mass of log$_{10}$(M$_{\rm BH}$) = 7.7$\pm$0.4 M$_{\odot}$ using the M--$\sigma$ relation of \citet{Mcconnell13}. 

AT2018fyk remained X-ray and UV bright for at least 500 days after discovery. Its properties (in particular the UV to X-ray spectral index \alphaox, X-ray spectrum and X-ray timing properties) showed similarities to outbursting stellar-mass black holes \citep{Wevers20, Wevers2021}, including the equivalent of the high/soft state (relatively UV bright SED, a weak non-thermal component in the X-ray spectrum, and a lack of high-frequency/short timescale/tens of minutes X-ray variability) and an accretion state transition into a low/hard state (relatively X-ray bright SED, non-thermal dominated X-ray spectrum and rapid X-ray variability on timescales of 1000s of seconds). \citet{Zhang22} reported soft X-ray time lags during this hard state (i.e., lower energy photons arrive later than higher energy photons); they found a lag of $\sim$1200 seconds in the 0.3--0.5 keV band (with respect to a reference band of 0.5--1 keV), decreasing monotonically with increasing energy. Around 600 days after discovery, the X-ray and the UV emission displayed a sudden and dramatic decrease (and an implied softening of the SED). We point out that the drop in X-ray and UV emission reveals a complete disconnect from the (expected) mass fall-back rate. Such lightcurve behavior (including rapid X-ray variability and a sudden decrease in luminosity) has not been seen in other sources, although it should be noted that the sample of X-ray TDEs (and in particular sources with similar observational coverage) is still very small. This was interpreted as the near-complete shutdown of accretion through a second state transition into quiescence or instability of the newly formed disk \citep{Wevers2021}.

\subsection{Rebrightening at very late times}
\begin{figure*}
    \centering
    \includegraphics[width=\textwidth]{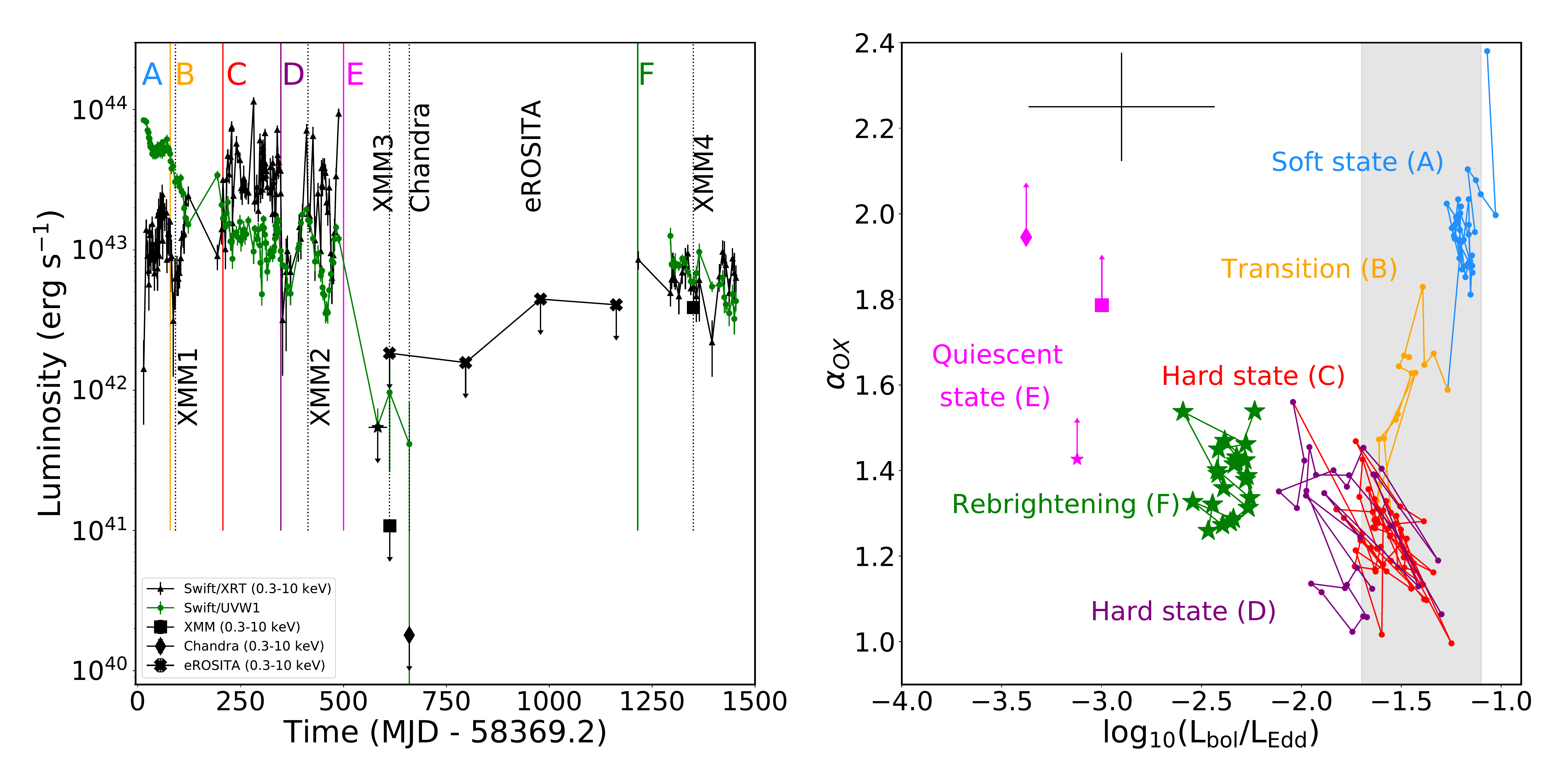}
    \includegraphics[width=0.48\textwidth]{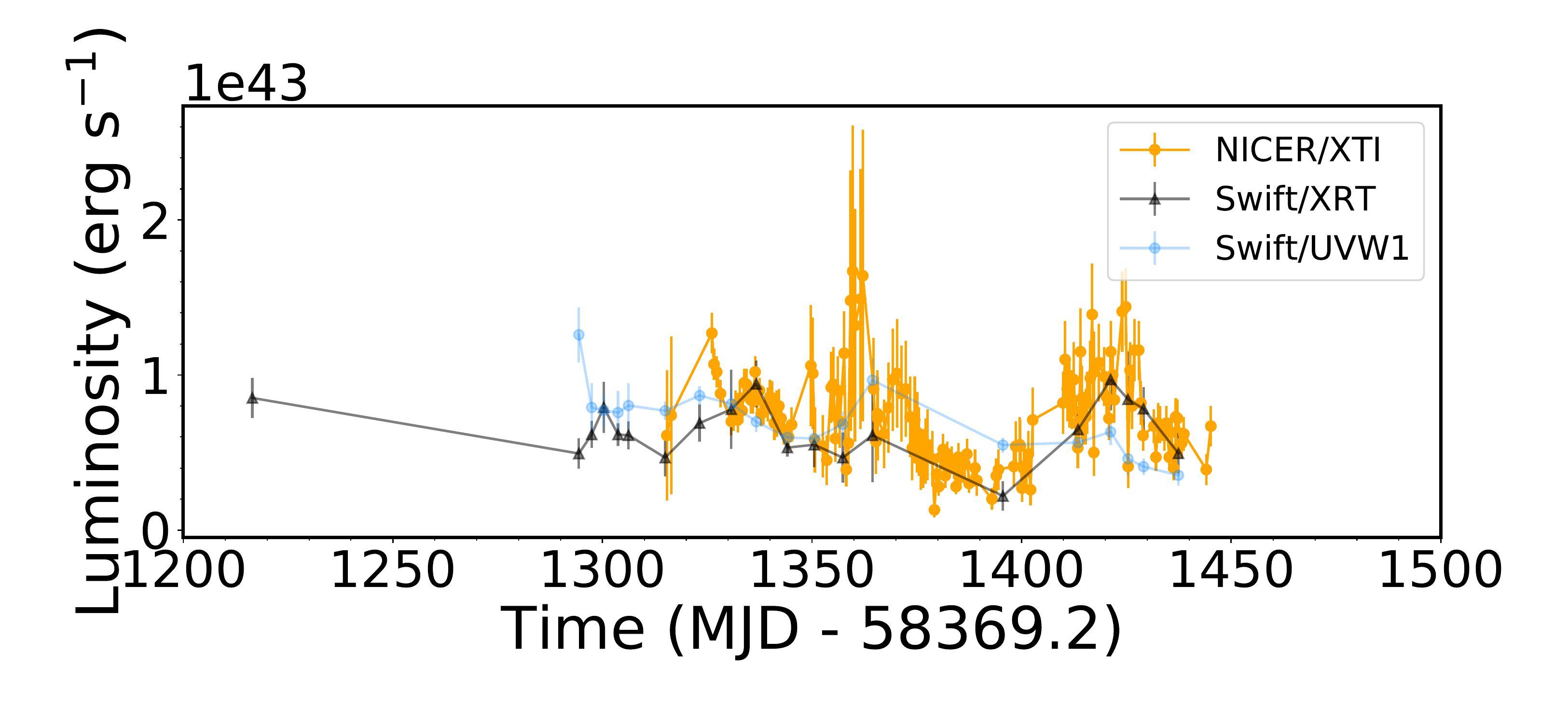}
    \includegraphics[width=0.48\textwidth]{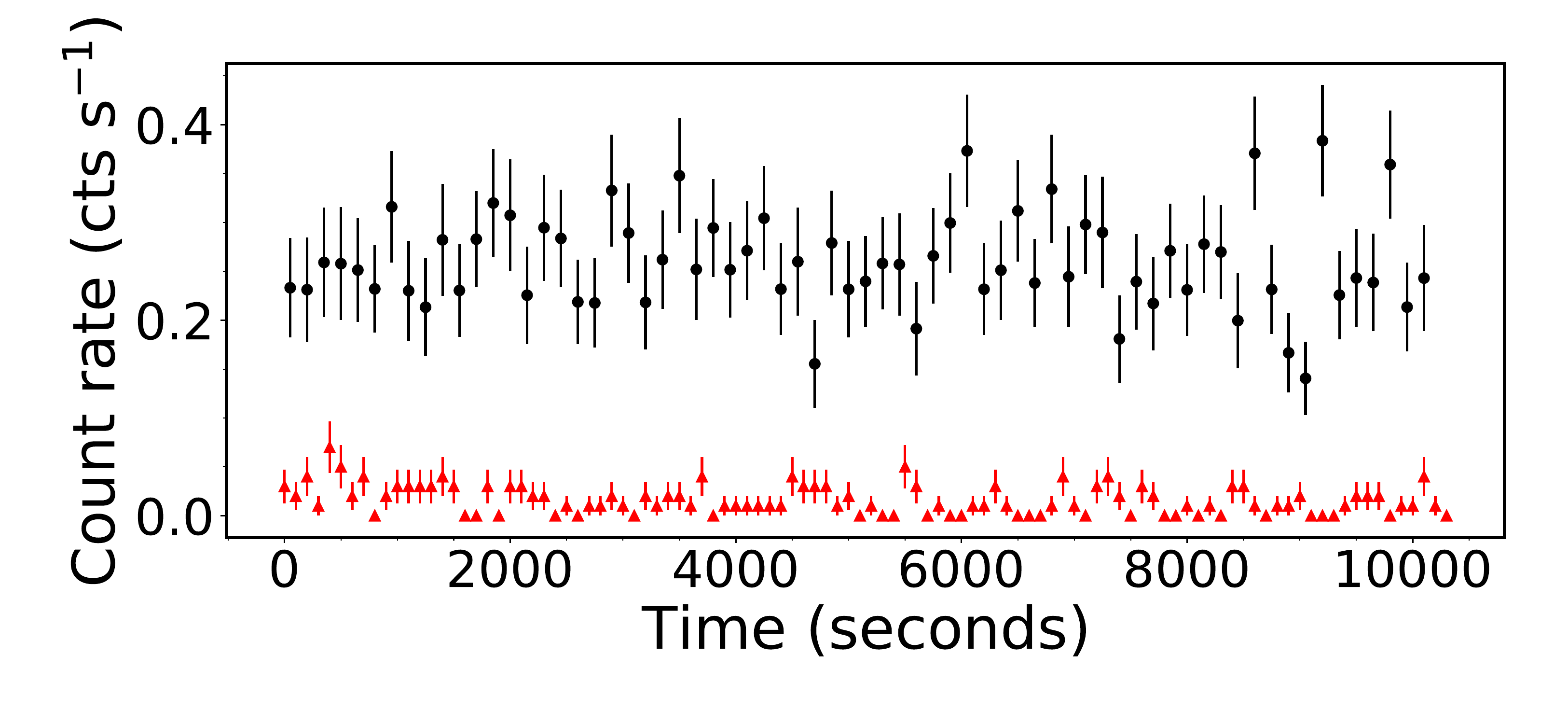}
    \caption{{\bf Top left}: Full X-ray (black, 0.3--10 keV) and \swift/UVW1 (green) lightcurve of AT2018fyk. Black crosses indicate \eros non-detections. Different source states are labelled A--F. {\bf Top right}: \alphaox vs. Eddington ratio, color coded by accretion state. The latest data are shown as green stars to distinguish them from the previous hard state. The observed behavior is consistent with a softening as the Eddington ratio decreases, which is also implied by the lower limits in the quiescent state. The black cross indicates the typical data uncertainty. {\bf Bottom:} \nicer/XTI lightcurve, overlaid on the \swift late-time data (left) and the XMM4 EPIC/PN (right) lightcurve (red triangles indicate the background rate). The \nicer light curve was extracted on a per GTI basis while the \xmm light curve has a binsize of 150 seconds. }
    \label{fig:alphaox}
\end{figure*}

Following the dramatic dimming after $\sim$600 days, seen at X-ray (by a factor $>6000$) and UV (by a factor $\sim$15) wavelengths, SRG/\eros \citep{Predehl21} scanned the position of AT2018fyk four times at phases of 611, 796, 978, and 1163 days after discovery. AT2018fyk was not detected in these epochs (see Figure \ref{fig:alphaox}), providing 3$\sigma$ upper limits (in the 0.3--2 keV band) of L$\rm _X \approx$ 1--3$\times$10$^{42}$ erg s$^{-1}$.

53 days after the last \eros non-detection, the source was detected again by {\it Neil Gehrels Swift} (\swift hereafter) monitoring observations obtained 1216 days after discovery with a luminosity of 8$\times$10$^{42}$ erg s$^{-1}$. This implies a relatively quick re-appearance of the X-ray emission. The X-ray brightness has increased by a factor of at least 100 compared to the deepest upper limit 700 days before and a factor 2--3 compared to the last \eros upper limit; the UV emission (0.03--3$\mu$m) has also brightened by a factor of $\approx$10 to L$_{\rm UV}$ = 7$\times$10$^{42}$ erg s$^{-1}$. 
Such behavior is both unprecedented and unexpected in the classical scenario of a star being fully disrupted by the SMBH. The data reduction for all new observations used in this work is described in the Supplementary Materials.

\subsubsection{Spectral energy distribution and X-ray spectrum}
\begin{figure}
    \centering
    \includegraphics[width=\textwidth]{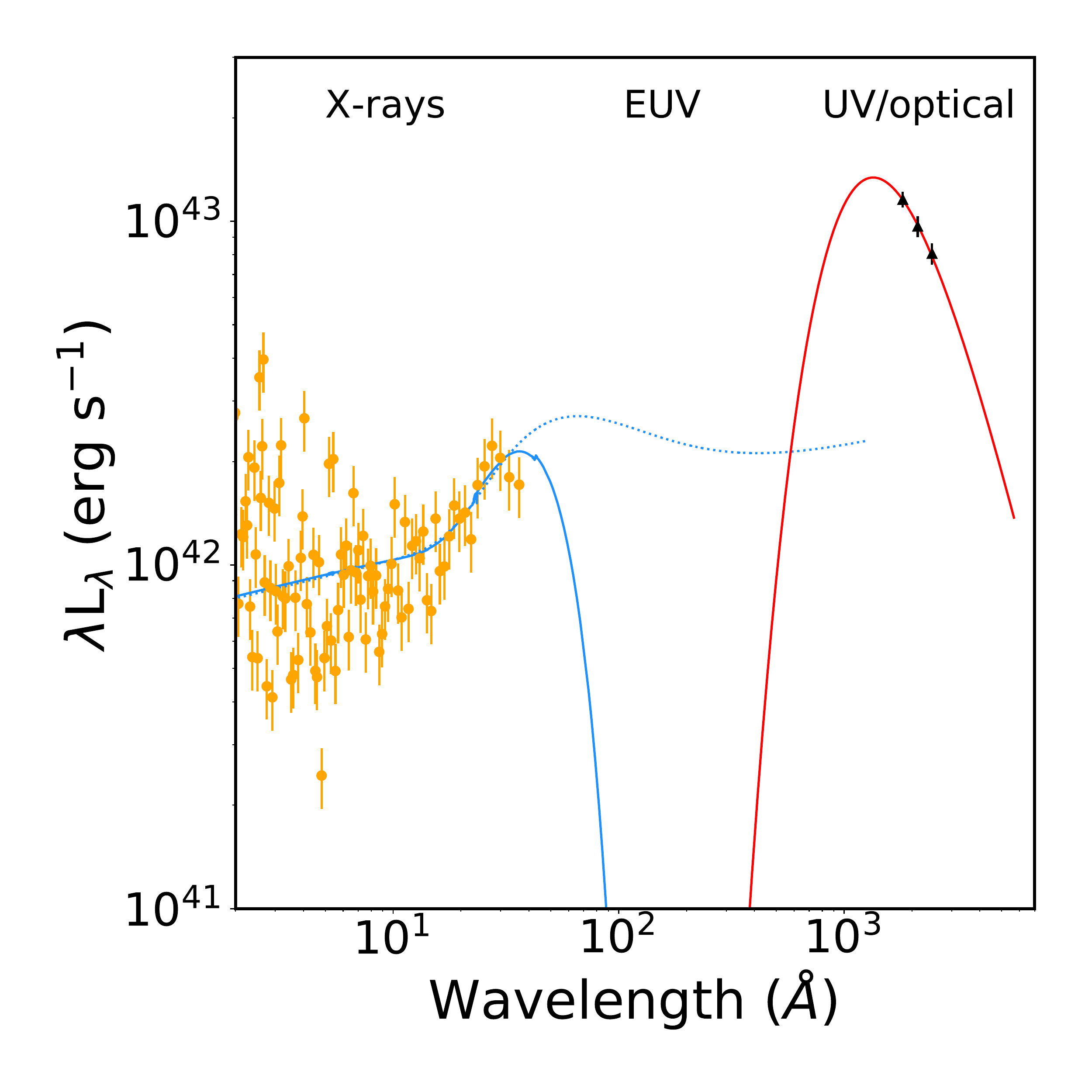}
    \caption{Observed UV/optical and X-ray spectral energy distribution at late times. XMM-Newton data are shown as orange circles, with the best-fit X-ray model (full line: absorbed, dotted line: unabsorbed) overplotted in blue. The UV/optical data are shown as black diamonds, with the red best-fit blackbody overlaid. Due to the lack of observational constraints in the EUV part of the SED, the bolometric emission (used to calculate the total radiated energy) is uncertain; for example, L$_{\rm bol}$ is calculated from the unabsorbed X-ray spectral model extrapolated to 10 eV ($\sim$1200 \AA).}
    \label{fig:sed}
\end{figure}

By modeling the available (host galaxy subtracted) \swift UV data with a blackbody function, we find a blackbody temperature of $\sim$25\,000 -- 35\,000 K, similar to the temperature at early times. We use this temperature to convert the UVW1 luminosity into the 0.03--3$\mu m$ emission\footnote{Note that the assumption of hot blackbody emission down to 0.03$\mu m$ cannot be verified, because the EUV is not observationally accessible (see also Figure \ref{fig:sed}).} (representing the total UV/optical emission, L$_{\rm UV}$).

We use XSPEC \citep{xspec} to model a new XMM-Newton (EPIC/PN) X-ray spectrum (obtained through director's discretionary time) with a phenomenological model ({\tt TBAbs $\times$ (diskbb + powerlaw)}) consisting of a thermal component and a power law, absorbed by a Galactic column of n$\rm _H$ = 1.15$\times$10$^{20}$ cm$^{-2}$ (i.e., the same model used in \citealt{Wevers2021}). We find a temperature of 113$\pm$31 eV for the thermal component, a power law index of $\Gamma$ = 2.2$\pm$0.2 and a power law fraction of emission (defined as the ratio of the power law flux to the total X-ray flux in the 0.3--10 keV band) of 80$\pm$10 per cent (full details are provided in Table \ref{tab:xrayspectra}). These values are all consistent with the previous hard state properties. Given these parameters, we calculate the conversion factor to first translate the count rate to 0.3--10 keV luminosity, and then convert this luminosity into the 0.01--10 keV luminosity (L$\rm _X$). 
We then calculate the Eddington fraction of emission as f$_{\rm Edd} = \frac{\rm L_{\rm UV} + L{\rm _X}}{\rm L_{\rm Edd}}$, that is, we assume that L$_{\rm bol}$ = L$_{\rm UV}$ + L$\rm _X$. 

Combining this power law fraction and index (consistent with the values obtained from \swift/XRT data) with the host subtracted and extinction corrected \swift/UVW1 fluxes, we calculate the UV to X-ray slope $\alpha_{\rm OX}$ of the late time emission. The full lightcurve and $\alpha_{\rm OX}$ as a function of bolometric Eddington ratio f$_{\rm Edd}$ is compared to the earlier evolution in Figure \ref{fig:alphaox} (top left and top right panels, respectively), while the late-time SED (including the XMM4 and Swift/UVOT data) is shown in Figure \ref{fig:sed}.  

We find that the spectral properties of AT2018fyk are very similar to those observed in the previous hard state observations (states C and D), just before the source became faint around day 500. The EPIC/PN lightcurve (Fig. \ref{fig:alphaox}, lower right panel) does not show statistically significant variability on timescales of 100--1000 seconds, although the uncertainties are large due to the relatively low count rate\footnote{There are hints of variability similar to that observed in the earlier hard state, but due to the larger error bars firm conclusions are not possible.}. When NICER restarted monitoring observations with a roughly twice daily cadence, several X-ray flaring episodes were observed (Fig. \ref{fig:alphaox}, lower left panel), which is not evident from the (lower cadence) \swift/XRT lightcurve. Significant variability on timescales of 6--12 hours is present throughout the \nicer observations. 

\subsection{Previous models for the X-ray and UV dimming}
Around day 500, the X-ray emission dropped by a factor $>6000$ in $\sim$170 days (from the Chandra observation; the XMM3 observation constrains the decrease to a factor $\sim$900 in $<$123 days), while the UV emission remained marginally detected above the host galaxy level, implying a drop by a factor of $\approx$15. The persistence of the UV emission implies a strong softening of the SED (measured through $\alpha_{OX}$, Figure \ref{fig:alphaox}, right panel) compared to the low/hard accretion state observed before the dimming event. 

Combined with the UV detections, the deep X-ray non detection followed by a re-detection might be due to the presence of a variable amount of optically thick material (e.g. neutral hydrogen). In order to explain the factor $\sim$6000 X-ray dimming, a column density of a few $\times 10^{24}$ cm$^{-2}$ is required. It seems unlikely that such a large ejection of material (e.g. in the form of a disk wind) would occur at the persistently low accretion rates ($\sim$0.1 of the Eddington rate, assuming a radiative efficiency $\eta$ = 0.1, see Fig. \ref{fig:alphaox}) that were observed. The sudden launching of such a disk wind 500 days after discovery would also be puzzling. The unbound debris provides an alternative (but equally unlikely) explanation. Assuming an outflow velocity of $\sim$10\,000 km s$^{-1}$, this material will span a large solid angle but will have diluted to densities $<< 10^{24}$ cm$^{-2}$; a variable obscuration model is also unlikely because this would imply that a single, lone cloud passed along our line of sight. 
High cadence X-ray and UV monitoring observations of AGNs similar to the data available for AT2018fyk show that most of these do not display significant flaring and/or dimming events (e.g. \citealt{Buisson2017}). Some of the most extreme AGNs have been observed to vary by a factor of at most several 100 (e.g. \citealt{Brandt1995, Forster1996, Boller2021}), highlighting that the observed behavior in AT2018fyk is atypical for AGNs.

Finally, \citet{Wevers2021} also explored the possibility of an accretion disk instability to explain the big drop in observed fluxes. Theoretical predictions suggest that the mass fall-back rate will evolve over time as $\dot{M} \propto t^{-5/3}$ (or even steeper, \citealt{Guillochon2013}), implying that the current mass fall-back rate should have decreased to lower levels ( $\sim$ 10$^{-1}$ of the peak $\dot{M}$, which corresponds to $\sim$ few $\times 0.01$ M$_{\odot}$ yr$^{-1}$). Furthermore, time-dependent TDE disk modeling suggests that even if this amount would be sufficient to {\it re-activate} the disk, it would then show short rebrightening bursts, rather than a sustained rebrightening at a steady luminosity \citep{Shen14}. We conclude that disk thermal instabilities are so poorly understood that they cannot be strongly ruled out, but we refrain from quantitatively considering them further.

\subsection{The host galaxy is not an AGN}\label{sec:hostagn}
\begin{figure}
    \centering
    \includegraphics[width=\textwidth]{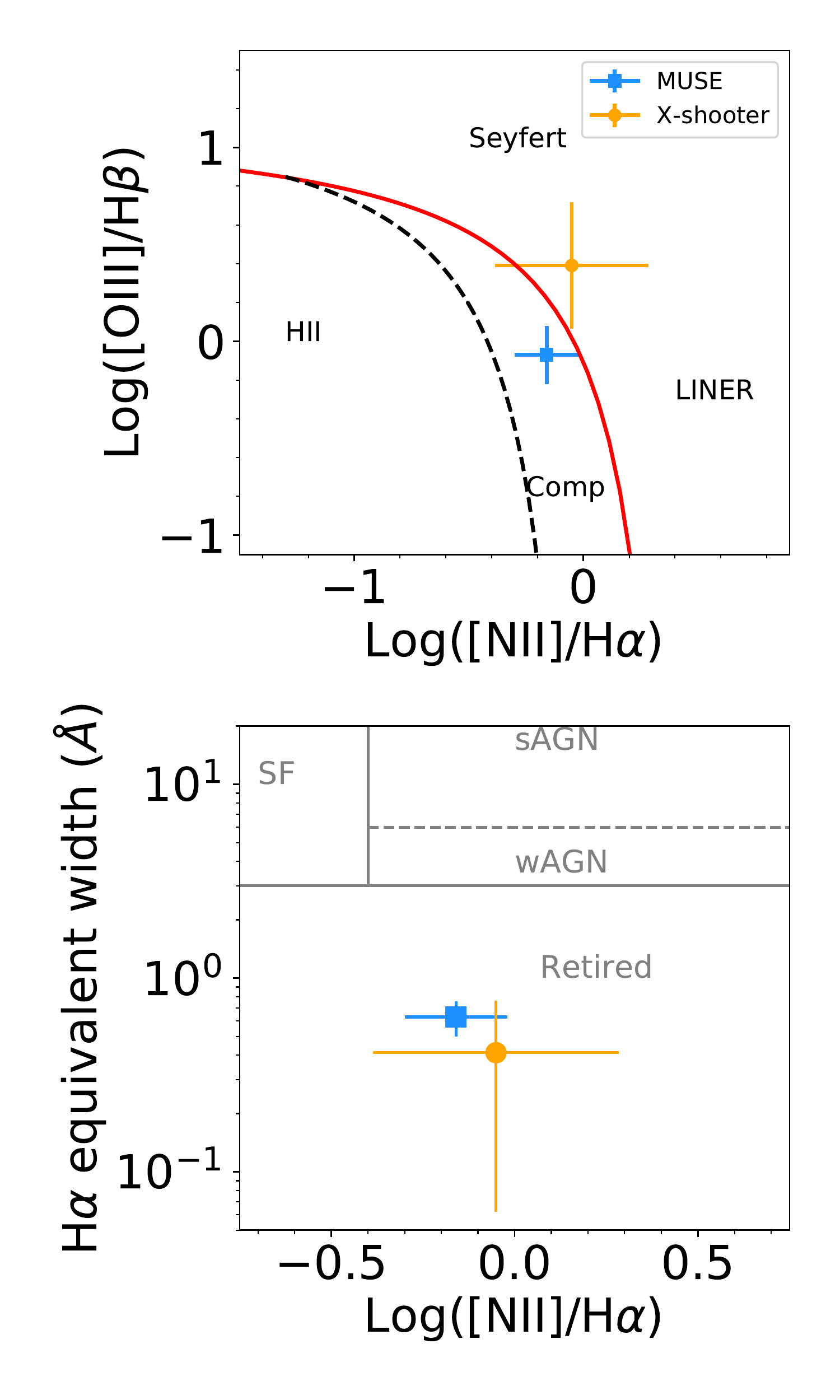}
    \caption{{\bf Top}: BPT diagram, with values measured from the template-subtracted spectrum. \target is located in the composite/LINER region. {\bf Bottom}: WHAN diagram, locating \target's host among the retired galaxy population.}
    \label{fig:bptwhan}
\end{figure}

In order to investigate the presence of an AGN, we inspect publicly available MUSE and X-shooter data (see Supplementary Materials) for emission lines. 
After modeling and subtracting the stellar continuum (see Figure \ref{fig:xshooter}), both the MUSE and the X-shooter data show very weak (EW$\sim$0.5--1 \AA) emission lines. We measured the line strengths and ratios to produce the BPT \citep{Baldwin1981} diagnostic diagram. The host galaxy is located in the Low-ionization nuclear emission-line region (LINER) region (Figure \ref{fig:bptwhan}). First thought to be exclusively produced by weak AGN \cite[e.g.,][]{Heckman1980,Kewley2006}, other ionisation mechanisms (unrelated to accretion) can also produce consistent line ratios \citep{Stasinska2008}. The WHAN diagram \citep{Cid2011} can differentiate a weak AGN from a retired galaxy\footnote{A retired galaxy has neither current star formation nor an active nucleus; instead, its post asymptotic giant branch stellar population can ionise the diffuse gas, producing EW H$\alpha$ up to 3 \AA, with line ratios that can occupy the LINER section of the BPT; see e.g. \citet[][]{Stasinska2008,Cid2010,Cid2011} for detailed discussions.} by substituting the [O\,\textsc{iii}]/H$\beta$ ratio for the equivalent width (EW) of H$\alpha$. Figure \ref{fig:bptwhan} shows the classification using AT2018fyk's host as a retired galaxy by the WHAN diagram. The lack of an increase in the H$\alpha$ EW towards the nucleus (see Fig. \ref{fig:muse}) also supports the non-AGN scenario.

To further investigate this interpretation, we also look at the infrared data: i) prior to 2018, the WISE \citep{Wright2010} IR W1 - W2 color of the host galaxy is $\approx$ 0.05, inconsistent with IR AGN selection criteria \citep[e.g.][]{Stern2012}; ii) \citet{Jiang2021} have analyzed the neoWISE IR light curve of AT2018fyk, and measured a covering factor ($f_c$)\footnote{The covering factor $f_c$ is defined as the ratio between the dust IR luminosity and the optical luminosity. It measures the fraction of the produced radiation that is absorbed by dust, hence the amount of dust in the nuclear region.} 
$\sim$0.01, indicating a dust/gas poor circumnuclear environment unlike those found in AGN \citep[$f_c > 0.3$,][]{Roseboom2013}.
 These results, consistent between datasets and wavelengths, provide the most robust evidence to date for the absence of an AGN in AT2018fyk. The implication is that AGN variability is strongly disfavored to explain the dramatic UV and X-ray variability seen in AT2018fyk.

\section{Explaining the rebrightening: a repeating partial tidal disruption event}
\label{sec:explanation}
\begin{figure*}
    \centering
    \includegraphics[width=\textwidth]{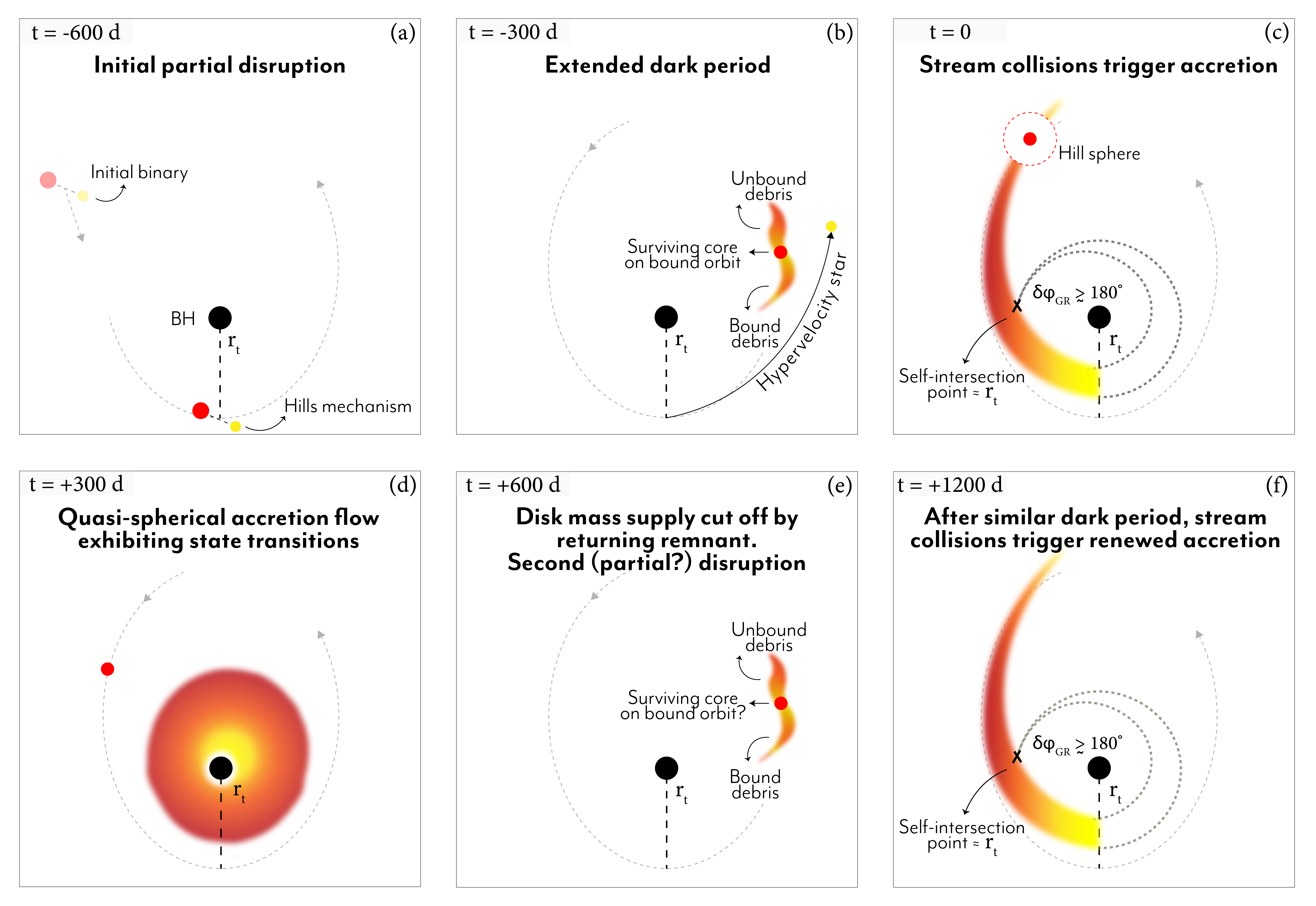}
    \caption{Multi-panel schematic indicating the various phases in the evolution of AT2018fyk. Time is indicated in the top left corner. The various components (SMBH, star, accretion flow) are not to scale. Following panel f), future observations will show whether the returning core was fully disrupted, or if a third dimming and rebrightening cycle occurs.}
    \label{fig:schematic}
\end{figure*}

For a $10^{7.7}M_{\odot}$ SMBH,
at most (for maximal spin) $\lesssim 5\%$ of stars with masses and radii comparable to those of the Sun (or smaller) will enter within the tidal radius, be destroyed completely, and not swallowed whole \citep{Kesden12, Ryu2020, Coughlin22b}. The tidal radius is also highly relativistic, suggesting that -- even for partial TDEs -- disk formation will be prompt, which is consistent with the observed properties of AT2018fyk (e.g., the presence of low ionisation Fe\,\textsc{ii} lines in the optical spectrum, the persistent X-ray brightness at UV/optical peak, the thermal X-ray spectrum at early times, and its short timescale variability in the X-rays, 
\citealt{Wevers19b, Wevers2021}). 
These arguments suggest that the star that initially fueled the outburst from AT2018fyk, by virtue of producing an observable flare, was partially disrupted (most TDEs will result in unobservable direct captures for the high black hole mass; see also \citealt{Coughlin22b}). Typically tidally disrupted stars are on approximately parabolic orbits (e.g., \citealt{Merritt2013}), 
which begs the question of how a partial TDE could yield a rebrightening because, as noted by \citet{Cufari2022a}, tidal dissipation within the partially disrupted star yields a minimum orbital period of a few $\times 10^{3}$ years for a $10^{7.7}M_{\odot}$ SMBH (see their Equation 1). One can bind the partially disrupted star more tightly if the star was \emph{initially} part of a binary system that was destroyed through Hills capture \citep{Hills1988}. In this case, the orbital period one would expect for the captured star is \citep{Cufari22}

\begin{equation}
    T_{\rm orb} \simeq \frac{2\pi}{\sqrt{GM_{\star}}}\left(\frac{a_{\star}}{2}\right)^{3/2}\left(\frac{M}{M_{\star}}\right)^{1/2}, \label{TorbHills}
\end{equation}
where $a_{\star}$ is the binary semimajor axis and $M_{\star}$ is the mass of the primary. A schematic of the different phases of the repeated partial disruption scenario and the timescales involved is shown in Figure \ref{fig:schematic}. 

With a host galaxy velocity dispersion of $\sigma = 158$ km s$^{-1}$, the maximum separation that a binary can have and still survive in the galactic nucleus is $a_{\star} \lesssim GM_{\star}/(4\sigma^2) \simeq 0.01$ AU (e.g., \citealt{Hills1975, Gould1991, Quinlan1996, Yu2002}). With $a_{\star} = 0.01$ AU, $M_{\star} = 1M_{\odot}$, and $M = 10^{7.7}M_{\odot}$, Equation \eqref{TorbHills} gives $T_{\rm orb} \simeq 2.5$ yr. 
A dynamical exchange can therefore produce a star on an orbit about the SMBH with a period as short as $\sim$ {\it a few} years. For separations $\lesssim 0.01$ AU, the tidal disruption radius of the binary is comparable to the tidal disruption radius of the star (increased by stellar rotation and relativistic effects; \citealt{Golightly2019, Gafton2015, Gafton2019}), and a partial TDE will occur (Figure \ref{fig:schematic}, panels a and b). The tight required separation of the initial binary provides constraints on the maximum size of the stars, in this case $\lesssim 2$ R$_{\odot}$. Such systems would require either two low mass stars, or a main sequence -- compact object binary; the latter (with the main-sequence star captured) is favored in order to reproduce the overall energetics and timescales of the TDE, as we now discuss (see Section \ref{sec:implications} for additional motivation for this type of binary).

Upon being partially disrupted, the material returns to the SMBH on a timescale that is approximately \citep{Lacy1982, Rees1988}

\begin{equation}
    T_{\rm acc} \simeq \frac{2\pi}{\sqrt{GM_{\star}}}\left(\frac{R_{\star}}{2}\right)^{3/2}\left(\frac{M}{M_{\star}}\right)^{1/2}, \label{Tacc}
\end{equation}
and has a peak magnitude 

\begin{multline}
    L_{\rm acc} \simeq 1.5\times 10^{45}\left(\frac{\eta}{0.1}\right) \left(\frac{M_{\star}}{M_{\odot}}\right)^{2}\left(\frac{R_{\star}}{R_{\odot}}\right)^{-3/2} \\ \times \left(\frac{M}{10^{7}M_{\odot}}\right)^{-1/2} 
    \textrm{ erg s}^{-1}, \label{Lacc}
\end{multline}
though these are generally longer and lower, respectively, for partial disruptions (e.g., \citealt{Guillochon2013, Miles2020, Nixon2021}; see also Section \ref{sec:implications} below). The proportionality coefficient in Equation \eqref{Lacc} matches simulations that yield a peak accretion rate equal to Eddington for $M = 10^{7}M_{\odot}$, $M_{\star} = 1 M_{\odot}$ and a radiative efficiency $\eta = 0.1$ \citep{Wu2018}. Setting $M = 10^{7.7}M_{\odot}$ and taking solar-like values gives $T_{\rm acc} \simeq 0.8$ yr and $L_{\rm acc} \simeq 6.7\times 10^{44}$ erg s$^{-1}$ (Figure \ref{fig:schematic}, panel b). Partial TDEs typically rise, peak, and decay as $\propto t^{-9/4}$ \citep{Coughlin2019, Miles2020, Nixon2021}, but for a star on a bound orbit, the fallback rate plummets as the star returns to pericenter \citep{liu22}. The reason for this sharp decline in the fallback rate is that the stellar core has a Hill sphere -- an approximately spherical region within which the star's gravitational field dominates over that of the SMBH -- near to which the stream density is much smaller than that of the bulk of the stream (Figure \ref{fig:schematic}, panel e). This feature of the fallback rate can physically explain the rapid shutoff displayed in Figure \ref{fig:alphaox} at $\sim 600$ days. 

While it likely does not inhibit the formation of a disk, nodal precession -- assuming the SMBH has a modest spin -- is probably important for its subsequent evolution: over many orbits of the material in the innermost regions of the disk, nodal and apsidal precession, coupled to the (likely) large misalignment angle between the spin axis of the SMBH and the angular momentum of the gas, will cause fluid annuli to precess independently instead of conforming to a smooth, warped disk \citep{Nixon2012, Liska21}. The orbit of the returning star also precesses and leads to a time-dependent feeding angle of the flow, and thus the gas is likely morphologically complex and, we suggest, closer to spherically symmetric than in the form of a traditional disc (see also \citealt{Patra22}). If we assume that the returning debris stream is cylindrical with cross-sectional radius $\sim R_{\odot}$ and length $a \simeq \frac{a_{\star}}{2}(M/M_{\odot})^{2/3}$ \citep{Cufari22}, then taking $a_{\star} = 0.01$ AU, $M = 10^{7.7}M_{\odot}$, and with $0.05 \, M_{\star}$ contained in the stream (see Section \ref{sec:implications}),
\begin{equation}
    \rho \simeq \frac{0.05 M_{\odot}}{\pi R_{\odot}^2a_{\star}\left(M/M_{\odot}\right)^{2/3}} \simeq 3.2\times10^{-7} \textrm{ g cm}^{-3}. \label{rhoinner}
\end{equation}
Taking $v = 0.1 \, c$ as the speed of the material as it shocks (recall the pericenter is highly relativistic), the shocked-gas pressure is

\begin{equation}
    p \simeq \rho v^2 \simeq 2.9\times10^{12} \textrm{ erg cm}^{-3}.
\end{equation}
The fluid is radiation-pressure dominated with a temperature

\begin{equation}
    T \simeq \left(\frac{3p}{a}\right)^{1/4} \simeq 5.8\times 10^{6}\textrm{ K} \simeq 0.5 \textrm{ keV}. \label{Tkev}
\end{equation}
Equation \eqref{Tkev} represents the self-intersection temperature near the horizon. The gas expands roughly adiabatically from the self-intersection point (e.g., \citealt{Jiang2016}), which reduces the temperature and density. At early times the gas will be optically thick, the photosphere at large radii, and the peak emission at temperatures below Equation \eqref{Tkev}. However, as time advances, the fallback rate declines, the density drops due to the continued expansion of the gas, and the flow becomes more optically thin to reveal the hot, inner regions, thus providing a plausible interpretation of the late-time dominance of the X-ray emission.

\section{Implications and Predictions}
\label{sec:implications}
From \citet{Wevers19b} and the additional data obtained since then, the total amount of energy radiated is equivalent to $\approx 9\times10^{51}$ erg. This energy could be up to a factor of $\sim 5$ smaller. The integral under the observed SED yields a luminosity lower by a factor of $\sim$5 compared to that of the total model SED (which yields the bolometric luminosity). The true value will be somewhere in between these two estimates. if the majority of the energy is not radiated at UV/optical wavelengths, as we have assumed in calculating the bolometric luminosity (and thus the total radiated energy; see Figure \ref{fig:sed}). If we adopt a radiative efficiency of $\eta = 0.1$, the radiated energy amounts to $\approx 0.05\,M_{\odot}$ of accreted mass. In normal TDEs (i.e., where the center of mass is on a parabolic orbit), approximately half of the stellar mass is accreted, and this implies that the star lost at most $\sim 0.1\,M_{\odot}$ during the tidal encounter. We note that for typical binaries the ratio of the binding energy of the binary to that of its stellar constituents is very small (on the order of the ratio of the stellar radius to the binary separation), and hence the approximation that only half of the material is accreted is usually warranted. Here, however, the binary must be very tight to reproduce the observed timescales, meaning that the binding energy of the binary is not substantially smaller than that of the star itself, and the ``unbound debris'' featured in panel b of Figure \ref{fig:schematic} may actually remain bound to the SMBH. If this is the case, we expect the luminosity of the ``lesser bound'' tail to be significantly lower than that of the more tightly bound tail owing to the longer return time. This material may be of sufficiently low density that it is substantially affected/destroyed by interactions with circumnuclear gas \citep{bonnerot16b} and the surviving core as it passes through pericenter a second time. Additional and more detailed investigations are required to constrain the energetics of the unbound/less bound tail.

Because the surviving core is spun up to near its breakup velocity, the tidal radius moves out \citep{Golightly2019}, and it is possible that the star was completely destroyed on its second pericenter passage (Figure \ref{fig:schematic}, panel f). If the mass lost from the star is closer to the upper limit of $\sim 0.1M_{\odot}$ that is inferred from the bolometric luminosity, then it could be that the star was completely destroyed on the second passage, and we would expect the accretion rate to monotonically decline with time. On the other hand, if the bolometric inference significantly overestimates the energy radiated and the mass accreted is closer to $\sim 0.01M_{\odot}$, it is likely that the star survived and will return to cause another dimming and future flare. Future observations will show if the star survived the second encounter to generate a third flare. 

From the \eros non-detections between days $\sim 600$ and $\sim 1200$ (although note that the last \eros upper limit is only a factor of $\sim$2--3 below the observed flux level), the fallback time of the material tidally stripped from the star during its second pericenter passage is, from Figure \ref{fig:alphaox}, $\sim 600$ days (note that ordinarily this timescale is virtually impossible to constrain from observations of single TDEs, but was possible because of fortuitous \eros data points). As noted above, the canonical timescale for a TDE between a solar-like star and a $10^{7.7}M_{\odot}$ SMBH is $T_{\rm acc} \simeq 0.8$ yr $\simeq 300$ days (see Equation \eqref{Tacc} above), which is a factor of $\sim 2$ shorter than the observed fallback time. 

However, the return time of the most-bound debris from a partial TDE can be significantly longer than the canonical value because of the gravitational influence of the surviving core, which is obvious from the fact that the fallback time becomes infinitely long in the limit that no mass is lost. From Figure 4 of \citet{Nixon2021}, the return time of the most bound debris from a $1M_{\odot}$, ZAMS star increases by a factor of $\sim 2-3$ in going from $\beta \simeq 2$ (where the disruption is full) to $\beta \simeq 1$ where the star loses $\sim 10\%$ of its mass (see Figure 4 of \citealt{Guillochon2013} and Figure 5 of \citealt{Mainetti2017}). \citet{Nixon2021} showed (top panel of their figure 2) that if the peak in the return time is extended to $\sim 0.2$ yr, implying a return time of $\sim 0.2$ yr $\times10^{1.7/2} \sim 520$ days for a $10^{7.7}M_{\odot}$ SMBH (comparable to what is observed for AT2018fyk), then we would need $\beta \lesssim 0.7$ if the star is somewhat evolved, and conceivably smaller if the star is near zero-age (Figure 1 of the same paper). 

For $\beta \simeq 0.7$, Figure 4 of \citet{Nixon2021} predicts a peak luminosity of $\sim 4\times10^{43}$ erg s$^{-1}$ (adopting a radiative efficiency $\eta = 0.1$) 
for a $10^{7.7}M_{\odot}$ SMBH, which is slightly less than but still in rough agreement with the X-ray luminosity in the top-left panel of Figure \ref{fig:alphaox}. At this value of $\beta$ the amount of mass lost from the star is also predicted from Newtonian simulations to be $\lesssim 0.01M_{\odot}$ (e.g., \citealt{Guillochon2013, lawsmith20}), which is in tension with the estimates from the bolometric luminosity that give a value that is closer to $\sim 0.1 M_{\odot}$. Nonetheless, as we noted the bolometric luminosity (and thus the total energy radiated) is uncertain for this system as it is based on a classic accretion disc model where the bulk of the energy is emitted at wavelengths for which we have no data, and the picture outlined here and shown in Figure \ref{fig:schematic} is clearly quite distinct from a standard disc; the total energy radiated could thus be a factor of $\sim 5$ smaller than the value used to infer the estimate of $0.05M_{\odot}$ accreted (see the discussion at the beginning of this section). Furthermore, general relativistic simulations indicate that more mass is lost from the star for the same $\beta$ as compared to Newtonian estimates; for example, Figure 3 of \citealt{Gafton2015}  shows that a $\beta = 0.7$ encounter between a solar-like $5/3$ polytrope and a $4\times 10^{7}M_{\odot}$ SMBH -- for which the pericenter distance is $\sim 5.8 GM/c^2$ -- strips $\sim 70\%$ of the stellar mass, while a $10^6M_{\odot}$ SMBH removes only $\sim 25\%$ for the same $\beta$. Figures 8 -- 12 of \citet{Gafton2019} show, nonetheless, that the timescales of the TDE remain similar. For our case in which a sun-like star disrupted by a $10^{7.7}M_{\odot}$ SMBH the tidal radius is $r_{\rm t} \simeq 3.5GM/c^2$ and thus highly relativistic. Hence even for $\beta \lesssim 0.7$ we would expect a larger fraction of the mass to be lost than would be predicted in the Newtonian limit. Thus, while more detailed modeling is required to more accurately constrain the properties of, e.g., the disrupted star, we find that the overall duration and energetics of the flare are consistent with the partial disruption of a near-solar star. On the other hand, increasing the mass and size of the star would increase the timescale, luminosity, and accreted mass and thus reduces these tensions, but the small separation of the binary restricts the size of the star to $\lesssim 1-2 R_{\odot}$ to avoid a common envelope phase (see also the last paragraph of this section).

Assuming that the first detection was approximately coincident with the time of the initial outburst, which is consistent with the lack of optical variability (e.g. from the pre-peak ASAS-SN lightcurve), we infer that the orbital period of the star is $\sim 1200$ days $\sim 3.3$ yr (i.e., the star's first pericenter passage was at day $\sim -600$ relative to discovery). We therefore predict that -- if the star was not destroyed on its second pericenter passage -- the source will abruptly decline in luminosity again around day 1800 (August 2023), before flaring for a third time (presuming the star is not destroyed on its third pericenter passage) around\footnote{The rapid rotation of the surviving core shortens the fallback time of the debris \citep{Golightly2019}, but we expect $\sim 2400$ days to roughly correspond with when the source will brighten a third time.} day $\sim 2400$ (March 2025). 

Finally, if the orbital period of the captured star is $\approx 1200$ days, then Equation \eqref{TorbHills} with $M_{\star} = 1M_{\odot}$ and $M/M_{\star} = 10^{7.7}$ suggests that the separation of the initial binary -- which was ripped apart to yield the captured star -- had a separation of $\sim 0.012$ AU. As noted above, the $M$-$\sigma$ relationship with a black hole mass of $10^{7.7}M_{\odot}$ implies that binaries must have a separation of less than $\sim 0.01$ AU to survive, and hence this binary separation is consistent with the high velocity dispersion in the nucleus of the galaxy. The distributions of observed binaries that are near solar are roughly uniform in semimajor axis or, for higher-mass stars, are uniform in $\log(a)$ (Opik's law) and thus peaked toward small separations \citep{offner22}. Since the hardening rate is roughly constant once the binary has reached a hardened separation \citep{Quinlan1996}, from a probabilistic standpoint we would also expect those with the widest (but hardened) initial separations to survive long enough to be fed into the galactic nucleus and tidally destroyed.

From the timescales and energetics arguments above (see the discussion around Equations \ref{Tacc} and \ref{Lacc}), the captured star that is repeatedly partially disrupted likely must be near-solar in terms of its mass and size. With a separation $a \lesssim 0.01$ AU $\sim 2R_{\odot}$, the companion object -- which was ejected during the separation of the binary (see panels a and b of Figure \ref{fig:schematic}) -- is therefore likely required to be a compact object to avoid being in a common envelope phase (as also argued in \citealt{Cufari22} in the context of the event ASASSN-14ko). If the companion was a white dwarf, which is most likely from a statistical standpoint, then the small binary separation appears consistent with the substantial population of detached white dwarf-main sequence binaries with semimajor axes $\sim 1R_{\odot}$ (likely as a consequence of a previous common envelope phase; e.g., \citealt{willems04, parsons15, mu21, hernandez21, zheng22}). Thus, in addition to being required from a survivability standpoint and to reproduce the orbital period of the captured star, the small separation of the binary is consistent if the companion is a white dwarf.

\section{Summary and conclusions}
\label{sec:summary}
After $\sim 600$ days of quiescence, the TDE AT2018fyk showed an anomalous rebrightening in both the UV and X-ray bands to luminosities to within a factor 10 of their peak values -- a behavior that is unprecedented in observations of TDEs. The model we propose to explain this behavior is that the initial flare was caused by the partial disruption of a star that was part of a binary system. The partially disrupted star was captured onto a relatively tight orbit through the destruction of the binary (i.e., Hills capture), thus generating a repeating, partial TDE (as well as a high velocity star flung out from the system) and the late-time flare. This model is not only consistent with the observations, but also predicts that 1) the fallback time of the tidally stripped debris is $\sim 600$ days (a timescale that is, we note, ordinarily very hard to constrain from observations of full TDEs), 2) the orbital time of the captured star is $\sim 1200$ days, and 3) the source should once again dim at day $\sim 1800$ (when the core is expected to return again) and brighten a third time at day $\sim 2400$ if the star was not completely destroyed on its second pericenter passage; on the other hand, if it was completely destroyed, we would expect -- as it is then an ordinary TDE -- a roughly power-law decay in its luminosity (although, if the star is on a bound orbit, it may exhibit a double-peaked lightcurve depending on the eccentricity; \citealt{Cufari2022a}).

We briefly remark that qualitatively similar behavior, including a late-time rebrightening in the background X-ray emission to $\sim 60\%$ of its peak magnitude around day $\sim 3600$ post-discovery, has recently been observed in a source exhibiting quasi-periodic X-ray eruptions (QPEs; \citealt{Miniutti22}). QPEs have also been hypothesised to be the result of repeated tidal stripping, particularly of white dwarfs by low mass SMBHs (e.g. \citealt{Arcodia21, King20, Miniutti22}), to explain their properties. Their host galaxies share several peculiar properties with those of TDEs, including low mass black holes and a preference for post-starburst galaxies \citep{Wevers22}. 

We finish by highlighting the importance of X-ray and UV monitoring observations of TDEs at late times. Almost all TDEs identified so far lack long-term (years long) follow-up. This leaves significant uncertainty as to whether similar behavior has occurred in other TDEs. For example, \citet{vanvelzen2019} report a deep UV upper limit for the source SDSS-TDE1, but no other meaningful constraints exist in the $\sim$6 years prior to that observation. Similarly, the majority of TDEs have either no or extremely sparse UV and X-ray constraints at late times. One exception to this is the recently reported observations of AT2021ehb, a TDE that similarly shows accretion state transitions at late times \citep{Yao2022}, although a partial TDE scenario is not necessary to explain that behavior. Long-term monitoring observations of TDEs -- particularly for those with high mass SMBHs where partial TDEs are very likely -- may provide more evidence for partial TDEs in the future. Indeed, highly periodic flaring may be among the most unambiguous signatures of a partial TDE in general.

\begin{acknowledgements}
We thank the anonymous referee for thoughtful comments and suggestions that helped to improve the manuscript.
We thank the \xmm, \swift, \nicer PIs (Norbert Schartel, Bradley Cenko and Keith Gendreau) and their operations teams for approving and promptly scheduling the requested observations. We warmly thank M.I. Saladino for help in creating Figure \ref{fig:schematic}. ERC thanks Chris Nixon for useful discussions, and acknowledges support from the National Science Foundation through grant AST-2006684 and from the Oakridge Associated Universities through a Ralph E.~Powe Junior Faculty Enhancement Award. Raw optical/UV/X-ray observations are available in the NASA/\swift archive (\url{http://heasarc.nasa.gov/docs/swift/archive}, Target Names: AT2018fyk, ASASSN-18UL); the XMM-Newton Science Archive (\url{http://nxsa.esac.esa.int}, obsID: 0911790601, 0911791601); \nicer data is publicly available through the HEASARC: \url{https://heasarc.gsfc.nasa.gov/cgi-bin/W3Browse/w3browse.pl}. Based on observations collected at the European Southern Observatory under ESO programmes 0103.D-0440(B) and 0106.21SS.

This work is based on data from eROSITA, the soft X-ray instrument aboard SRG, a joint Russian-German science mission supported by the Russian Space Agency (Roskosmos), in the interests of the Russian Academy of Sciences represented by its Space Research Institute (IKI), and the Deutsches Zentrum für Luft- und Raumfahrt (DLR). The SRG spacecraft was built by Lavochkin Association (NPOL) and its subcontractors, and is operated by NPOL with support from the Max Planck Institute for Extraterrestrial Physics (MPE).

The development and construction of the eROSITA X-ray instrument was led by MPE, with contributions from the Dr. Karl Remeis Observatory Bamberg \& ECAP (FAU Erlangen-Nuernberg), the University of Hamburg Observatory, the Leibniz Institute for Astrophysics Potsdam (AIP), and the Institute for Astronomy and Astrophysics of the University of Tübingen, with the support of DLR and the Max Planck Society. The Argelander Institute for Astronomy of the University of Bonn and the Ludwig Maximilians Universität Munich also participated in the science preparation for eROSITA.

The eROSITA data shown here were processed using the eSASS software system developed by the German eROSITA consortium.
\end{acknowledgements}
\newpage
\appendix
\setcounter{table}{0}
\renewcommand{\thetable}{A\arabic{table}}
\setcounter{figure}{0}
\renewcommand{\thefigure}{A\arabic{figure}}

\section*{Supplementary material}
\section{Observations and data reduction}
\label{sec:observations}
\subsection{\swift XRT and UVOT}
We reduce the UV/Optical Telescope (UVOT; \citealt{Roming05}) data using the {\tt uvotsource} task, extracting fluxes from the standard 5 arcsec aperture. We subsequently correct for Galactic extinction assuming E(B-V) = 0.01 \citep{Schlafly2011}, and subtract the host galaxy contribution as determined from SED fitting in \citet{Wevers2021}. The emission in the UV bands has brightened by a factor of $\sim$10, although in the optical this is much less pronounced with the brightness in the B and V filters remaining consistent with the inferred host galaxy brightness. We therefore do not include these filters in our analysis. The UV lightcurves can be found in the online supplementary material. The \swift/XRT lightcurve and late time stacked spectrum were extracted using the online XRT tool\footnote{https://www.swift.ac.uk/user$\_$objects/}. 

\subsection{XMM-Newton}
\label{sec:xmm}
A 29 kilosecond observation was approved by the XMM-Newton director and executed on 2022 May 20/21 (obsid: 0911790601). The optical monitor used the UVW1 filter, taking 5 deep images as well as a small window centred on the galaxy nucleus with data in time-tag (fast) mode. The EPIC instruments (PN, MOS1 and MOS2) were operated in full frame mode with the {\it thin1} filter. The observation was split into 2 blocks, one of 20 ks and one of 9 ks. The latter was unfortunately lost due to telemetry problems. An additional 10 ks observation was therefore scheduled on 2022 June 9, with an identical instrument setup (obsid: 0911791401).

We start by reprocessing the data using the {\tt emproc} and {\tt epproc} tasks in XMM-SAS v1.3. Good time intervals are identified by excluding periods of background flaring in the 10--12 keV band. This leaves approximately 9.2 ks of exposure for the observation with ID 0911790601, while 3.5 ks remains for ID 0911791401. We therefore only use the data of obsID 0911790601 for our analysis. The background is estimated from a source-free region with radius 50 arcsec on the same detector, while the source signal is extracted from a region with radius 33 arcsec. After applying standard data filters, we extract spectra and lightcurves in the 0.3--10 keV energy range. Lightcurves are further corrected for instrumental effects using the {\tt epiclccorr} task.

\begin{table*}
\caption{Best fit parameters obtained from X-ray spectral modeling of the (stacked) \swift and \xmm data. The mean count rate for each spectrum is given in the second column. The effective exposure time t$_{\rm exp}$ is given in kiloseconds. The spectral model used is {\tt TBabs*zashift*(diskbb + powerlaw)}; kT is the temperature of the thermal component, while $\Gamma$ denotes the power-law spectral index. Normalizations for the thermal and power-law components are listed in the norm(kT) and norm($\Gamma$) columns. The flux is integrated from 0.3--10 keV. PL frac denotes the fractional contribution of the power-law component to the total X-ray flux. The final column lists the reduced $\chi^2$ and degrees of freedom (dof).}
    \centering
    \begin{tabular}{ccccccccccc}\hline\hline
    Spectrum & Count rate & State & t$_{\rm exp}$ & kT & norm(kT) & $\Gamma$ & log$_{10}$(norm ($\Gamma$)) & log$_{10}$(flux) & PL frac & $\chi^2$ (dof) \\\hline
XRT & 0.011 & F & 45650 & 175$\pm$60 & 8$^{+42}_{-6}$ & 2.15$\pm$0.4 & --4.2$\pm$0.2 & --12.35$\pm$0.04 & 79$\pm$10 & 23 (24) \\ 
PN (0601) & 0.25 & F & 9200 & 113$\pm$31 & 102$^{+458}_{-71}$ & 2.16$\pm$0.2 & --4.10 $\pm$0.08 & --12.37$\pm$0.03 & 80$\pm$10 & 114 (113)\\
PN (1401) & 0.30 & F & 3500 & 152$\pm$80 & 24$^{+900}_{-20}$ & 2.4$\pm$1.4 & --4.07 $\pm$0.5 & --12.24$\pm$0.2 & 80$\pm$10 & 203 (170)\\\hline    
\end{tabular}
\label{tab:xrayspectra}
\end{table*}

\subsection{\nicer/XTI}
\label{sec:nicer}
\nicer is a non-imaging detector with 52 co-aligned concentrators that focus X-rays onto silicon drift detectors at their respective foci. It has a field of view of 3.1 arcmins in radius and a nominal bandpass of 0.2--12 keV. But depending on the source brightness and background the usable bandpass can vary. \nicer's large effective area of $>$1700 cm$^{2}$ at 1 keV enabled by its 52 Focal Plane Modules (FPMs), ability to steer rapidly to any part of the sky, and monitor sources for extended periods of months and years makes it an excellent telescope for tracking long-term transients like TDEs.

Following the \swift/XRT detection of \target \nicer started a high-cadence monitoring program as part of an approved guest observer program (ID: 5070, PI: Pasham). \nicer data is organized in the form of obsids which represent a collection of short exposures or good time intervals (GTIs) varying between 100 s to upto 2000 s over the time span of a day. While \nicer monitoring of \target continues at the time of writing of this paper we include all data taken prior to 22 August 2022. 

We started our \nicer data analysis by downloading the raw, unfiltered (uf) data from the HEASARC public archive. These were reduced using the standard reduction procedures of running {\it nicerl2} followed by {\it nimaketime}. All the filter parameters except for {\tt overonly\_range}, {\tt underonly\_range}, and {\tt overonly\_expr}, were set to the default values as recommended by the data analysis guide: \url{https://heasarc.gsfc.nasa.gov/lheasoft/ftools/headas/nimaketime.html}. The reason for not screening on undershoots and overshoots is to ensure we are not throwing away good data in the name of strict default screening values. Instead we screen each GTI based on the net, i.e., background-subtracted, 0--0.2 keV, 13--15 keV, and 4--12 keV count rates as recommended by \cite{3c50}. After a background spectrum is estimated using the 3c50 model, if the absolute value of the net count rate in the 0--0.2 keV is more than 2 cps, if the absolute value of the net rate in 13--15 keV is more than 0.05 cps, or if the absolute value of the net 4--12 keV is more than 0.5 cps, we mark that GTI as bad and omit it from further analysis (see \citealt{pashcow} for more details).

To improve statistics, we also extracted 18 time-resolved spectra by combining multiple GTIs. Spectra were binned with the optimal binning scheme of \citep{optmin}. To do this we used the {\it ftool} {\it ftgrouppha} with an additional requirement to have a minimum of 20 counts per spectral bin. \target was above the background in the 0.3--0.7 keV bandpass. Because of this limited bandpass we fit each spectrum with a simple powerlaw plus a Gaussian model ({\it tbabs*zashift(pow) + gaussian in XSPEC}) and inferred the best-fit power-law index and absorption corrected 0.3--10 keV luminosities. The {\it gaussian} component was used to model out the variable strength background Oxygen line at 0.54 keV from the Earth's atmosphere. A summary of the spectral modeling is shown in Table \ref{tab:nicer}.\\

\ttabbox[\linewidth]{
\resizebox{\textwidth}{!}{\begin{tabular}{*{10}{l}}
\toprule
\toprule
\multicolumn{10}{c}{Best-fit parameters from fitting time-resolved 0.3-0.7 keV \nicer~ X-ray spectra} \\
\bottomrule
{\bf Start} & {\bf End} & {\bf Exposure} & {\bf FPMs} & {\bf Phase} & {\bf $\Gamma$} & {\bf Log(Integ. Lum.)} & {\bf Log(Obs. Lum.)} & {\bf Count rate } & {\bf Gaussian}  \\
 (MJD) & (MJD) & (ks) & & & & (0.3-10 keV) & (0.3-10.0 keV) & (0.3-10.0 keV) & norm \\
\midrule
59682.51 & 59689.0 & 0.87 & 51 & L1 & 2.64$^{+0.6}_{-0.53}$ & 42.8$^{+0.21}_{-0.16}$ & 42.7$^{+0.14}_{-0.19}$ & 0.0055$\pm$0.0024 & 1.0$^{+2.3}_{-1.0}$ \\
59693.7 & 59698.27 & 2.6 & 43 & L2 & 4.49$^{+0.24}_{-0.23}$ & 42.99$^{+0.02}_{-0.02}$ & 42.9$^{+0.01}_{-0.01}$ & 0.0166$\pm$0.0012 & 23.1$^{+1.7}_{-2.4}$ \\
59698.27 & 59703.0 & 5.45 & 47 & L3 & 3.47$^{+0.22}_{-0.21}$ & 42.84$^{+0.03}_{-0.03}$ & 42.74$^{+0.03}_{-0.02}$ & 0.0093$\pm$0.0006 & 7.7$^{+1.1}_{-1.2}$ \\
59703.0 & 59708.0 & 8.68 & 46 & L4 & 3.26$^{+0.18}_{-0.18}$ & 42.88$^{+0.03}_{-0.03}$ & 42.77$^{+0.03}_{-0.03}$ & 0.0095$\pm$0.0003 & 5.3$^{+1.0}_{-0.9}$\\
59708.0 & 59718.0 & 9.1 & 49 & L5 & 2.77$^{+0.17}_{-0.17}$ & 42.82$^{+0.05}_{-0.04}$ & 42.75$^{+0.04}_{-0.04}$ & 0.0073$\pm$0.0003 & 5.5$^{+0.8}_{-0.8}$ \\
59718.0 & 59723.0 & 3.46 & 51 & L6 & 2.53$^{+0.36}_{-0.35}$ & 42.79$^{+0.14}_{-0.11}$ & 42.72$^{+0.1}_{-0.12}$ & 0.0057$\pm$0.0007 & 3.9$^{+1.2}_{-1.2}$ \\
59723.0 & 59728.0 & 5.12 & 50 & L7 & 2.39$^{+0.22}_{-0.22}$ & 42.84$^{+0.09}_{-0.08}$ & 42.77$^{+0.06}_{-0.06}$ & 0.0055$\pm$0.0005 & 2.6$^{+0.8}_{-0.8}$ \\
59728.0 & 59733.0 & 3.69 & 50 & L8 & 1.81$^{+0.34}_{-0.33}$ & 43.12$^{+0.23}_{-0.19}$ & 43.06$^{+0.27}_{-0.24}$ & 0.0052$\pm$0.0007 & 0.7$^{+0.7}_{-0.7}$ \\
59733.0 & 59738.0 & 3.63 & 52 & L9 & 2.48$^{+0.36}_{-0.33}$ & 42.82$^{+0.14}_{-0.11}$ & 42.73$^{+0.15}_{-0.11}$ & 0.0056$\pm$0.0007 & 1.9$^{+1.1}_{-1.1}$ \\
59738.0 & 59743.0 & 3.23 & 52 & L10 & 2.33$^{+0.32}_{-0.29}$ & 42.91$^{+0.13}_{-0.11}$ & 42.83$^{+0.14}_{-0.16}$ & 0.0055$\pm$0.0008 & 0.0$^{+0.3}_{-0.0}$ \\
59743.0 & 59748.0 & 4.76 & 52 & L11 & 2.5$^{+0.38}_{-0.36}$ & 42.7$^{+0.15}_{-0.11}$ & 42.61$^{+0.16}_{-0.12}$ & 0.0041$\pm$0.0005 & 0.9$^{+0.9}_{-0.9}$ \\
59748.0 & 59758.0 & 10.3 & 52 & L12 & 2.84$^{+0.21}_{-0.22}$ & 42.55$^{+0.06}_{-0.04}$ & 42.46$^{+0.06}_{-0.05}$ & 0.0038$\pm$0.0002 & 1.9$^{+0.5}_{-0.5}$ \\
59758.0 & 59768.0 & 2.53 & 51 & L13 & 3.44$^{+0.69}_{-0.6}$ & 42.59$^{+0.1}_{-0.06}$ & 42.51$^{+0.07}_{-0.06}$ & 0.0056$\pm$0.0009 & 5.8$^{+1.5}_{-1.5}$ \\
59768.0 & 59778.0 & 4.14 & 51 & L14 & 2.68$^{+0.35}_{-0.34}$ & 42.62$^{+0.11}_{-0.09}$ & 42.57$^{+0.09}_{-0.07}$ & 0.0048$\pm$0.0006 & 5.8$^{+0.9}_{-1.0}$ \\
59778.0 & 59783.0 & 6.03 & 52 & L15 & 2.36$^{+0.21}_{-0.21}$ & 42.85$^{+0.09}_{-0.07}$ & 42.77$^{+0.11}_{-0.07}$ & 0.0055$\pm$0.0004 & 2.1$^{+0.6}_{-0.4}$ \\
59783.0 & 59788.0 & 5.06 & 52 & L16 & 2.38$^{+0.22}_{-0.21}$ & 42.92$^{+0.09}_{-0.08}$ & 42.83$^{+0.07}_{-0.07}$ & 0.006$\pm$0.0005 & 0.6$^{+0.4}_{-0.6}$ \\
59788.0 & 59798.0 & 7.05 & 52 & L17 & 2.37$^{+0.16}_{-0.16}$ & 42.96$^{+0.06}_{-0.05}$ & 42.87$^{+0.04}_{-0.05}$ & 0.0066$\pm$0.0004 & 0.0$^{+0.0}_{-0.0}$ \\
59798.0 & 59820.0 & 8.22 & 52 & L18 & 2.82$^{+0.2}_{-0.2}$ & 42.74$^{+0.05}_{-0.04}$ & 42.63$^{+0.06}_{-0.06}$ & 0.0052$\pm$0.0003 & 0.7$^{+0.7}_{-0.5}$ \\
 \bottomrule
\bottomrule
\end{tabular}}
 }
{\caption{{\bf Summary of time-resolved X-ray energy spectral modeling of \target}. Here, 0.3--0.7 keV \nicer spectra are fit with {\it tbabs*zashift(clumin*pow) + gaussian} model using {\it XSPEC} \citep{xspec}. {\bf Start} and {\bf End} represent the start and end times (in units of MJD) of the interval used to extract a combined \nicer spectrum. {\bf Exposure} is the accumulated exposure time during this time interval. {\bf FPMs:} The total number of active detectors minus the ``hot'' detectors. {\bf Phase} is the name used to identify the epoch. {\bf $\Gamma$} is the photon index of the powerlaw component. {\bf Log(Integ. Lum.)} is the logarithm of the integrated absorption-corrected powerlaw luminosity in 0.3-10 keV in units of erg s$^{-1}$. {\bf Log(Obs. Lum.)} is the logarithm of the observed, extrapolated 0.3--10.0 keV luminosity in units of erg s$^{-1}$. {\bf Count Rate } is the background-subtracted \nicer count rate in 0.3--0.7 keV in units of counts/sec/50 FPMs. All errorbars represent 1-$\sigma$ uncertainties. The total best-fit $\chi^{2}$/the degrees of freedom over all the spectra is 76.5/65. }\label{tab:nicer}}

\subsection{SRG/\eros}
Coinciding with the quiescent phase following the first major optical outburst, AT~2018fyk was observed every 6 months by \textit{SRG}/\eros \citep{Sunyaev21, Predehl21} during its first four all sky surveys (denoted eRASS1, 2, 3 and 4, respectively; a log of observations is presented in Table~\ref{tab:erosita_observation_log}). No X-ray point source was detected by the \eros Science Analysis Software pipeline (eSASS; \citealt{Brunner22}) within 60$^{\prime \prime}$ of the optical position of AT~2018fyk during these scans. Using the eSASS task SRCTOOL (v211214), source counts were extracted from a circular aperture of radius 30$^{\prime \prime}$ centered on the optical position of AT~2018fyk, whilst background counts were extracted from a source-free annulus with inner and outer radii 140$^{\prime \prime}$ and 240$^{\prime \prime}$, respectively. The inferred 3$\sigma$ upper limits on the 0.3--2~keV count rates in each eRASS scan were (0.067, 0.063, 0.16, 0.14)~cts~s$^{-1}$, on MJD (58981.348, 59165.818, 59348.557, and 59532.943), respectively. Assuming the best fitting spectral model from the \textit{XMM} observation in Table~1, then these rates correspond to upper limits on the 0.3--2~keV observed fluxes of (1.4, 1.2, 3.4, and 3.1)$\times 10^{-13}$~erg~s$^{-1}$~cm$^{-2}$, respectively. A 0.3--2~keV 3$\sigma$ upper limit on the source count rate from the stack of eRASS1-4 observations is 0.032~cts~s$^{-1}$ (observed 0.3--2~keV flux of $6.5\times 10^{-14}$~erg~s$^{-1}$~cm$^{-2}$). Based on the spectral model derived from the XMM-Newton observation, we calculate a correction factor of 1.46 for the conversion from the 0.3--2 keV to 0.3--10 keV band. We report the 0.3--10 keV band values throughout the manuscript for consistency with data from other observatories.

\begin{table*}[]
    \centering
    \begin{tabular}{c|c|c|c|c|c|c}
        eRASS & Exposure & MJD start & MJD stop & Phase & Rate & $F_{\mathrm{X, obs}}$\\
         & [s] & & & [days] & [cts s$^{-1}$] & [$10^{-13}$ erg~s$^{-1}$~cm$^{-2}$] \\
        \hline
        eRASS1 & 206 & 58980.848 & 58981.848 & 612.148 & $<$0.067 & $<$1.4\\
        eRASS2 & 172 & 59165.401 & 59166.235 & 796.618 & $<$0.063 & $<$1.2\\
        eRASS3 & 124 & 59348.223 & 59348.890 & 979.357 & $<$0.162 & $<$3.4\\
        eRASS4 & 179 & 59532.610 & 59533.277 & 1163.743 & $<$0.138 & $<$3.1\\
    \end{tabular}
    \caption{Log of \textit{SRG}/ \eros observations of AT~2018fyk during its all sky survey. The MJD start and stop columns refer to the times of the first and last observation of AT~2018fyk within a given eRASS, with the phase then being measured based on the midpoint of these relative to MJD=58369.2. The rate and $F_{\mathrm{X, obs}}$ columns are the observed source count rates and observed fluxes computed in the 0.3-2~keV band (not corrected for Galactic absorption).}
    \label{tab:erosita_observation_log}
\end{table*}

\subsection{MUSE}
\label{sec:muse}
AT2018fyk was observed by the Multi Unit Spectroscopic Explorer \cite[MUSE,][]{Bacon_2010} on 2019 June 10 (MJD 58644) as part of the All-weather MUse Supernova Integral field Nearby Galaxies (AMUSING) survey, ESO ID: 0103.D-0440(B). At this epoch, the transient was already quiescent at optical wavelengths, and the host galaxy emission completely dominated the data. The data cube was analysed as part of the AMUSING++ Nearby Galaxy Compilation \citep{Lopez-Coba_2020}. However, the authors did not include it in the final sample of the paper due to the lack of strong emission lines, which were the main subject of their study. Nevertheless, we obtained the final products of their stellar population and emission-line fitting analyses (López-Cobá, private communication). 

A detailed description is presented in \citet{Lopez-Coba_2020}. In summary, the following procedure was adopted. First, the raw data cubes were reduced with REFLEX \citep{Freudling_2013} using version 0.18.5 of the MUSE pipeline. Next, the emission lines and stellar population content were analysed using the PIPE3D pipeline \citep{Sanchez_2016a}, a fitting routine adapted to analyse IFS data using the package FIT3D \citep{Sanchez_2016b}. The procedure starts by performing a spatial binning on the continuum (V-band) to increase the signal-to-noise ratio in each spectrum of the data-cube. The stellar population model was derived by performing stellar population synthesis; the PIPE3D implementation adopts the GSD156 stellar library, which comprises 39 ages and four metallicities, extensively described in \citet{Cid_Fernandes_2013}.
Then, a model of the stellar continuum in each spaxel was recovered by re-scaling the model within each spatial bin to the continuum flux intensity in the corresponding spaxel. The best model for the continuum was then subtracted to create a pure gas data cube.
A set of 30 emission lines within the MUSE wavelength range were fit spaxel by spaxel for the pure gas cube by performing a non-parametric method based on a moment analysis. The data products of the pipeline are a set of bi-dimensional maps of the considered parameters with their corresponding errors. 

\begin{figure}
    \centering
    \includegraphics[width=\textwidth]{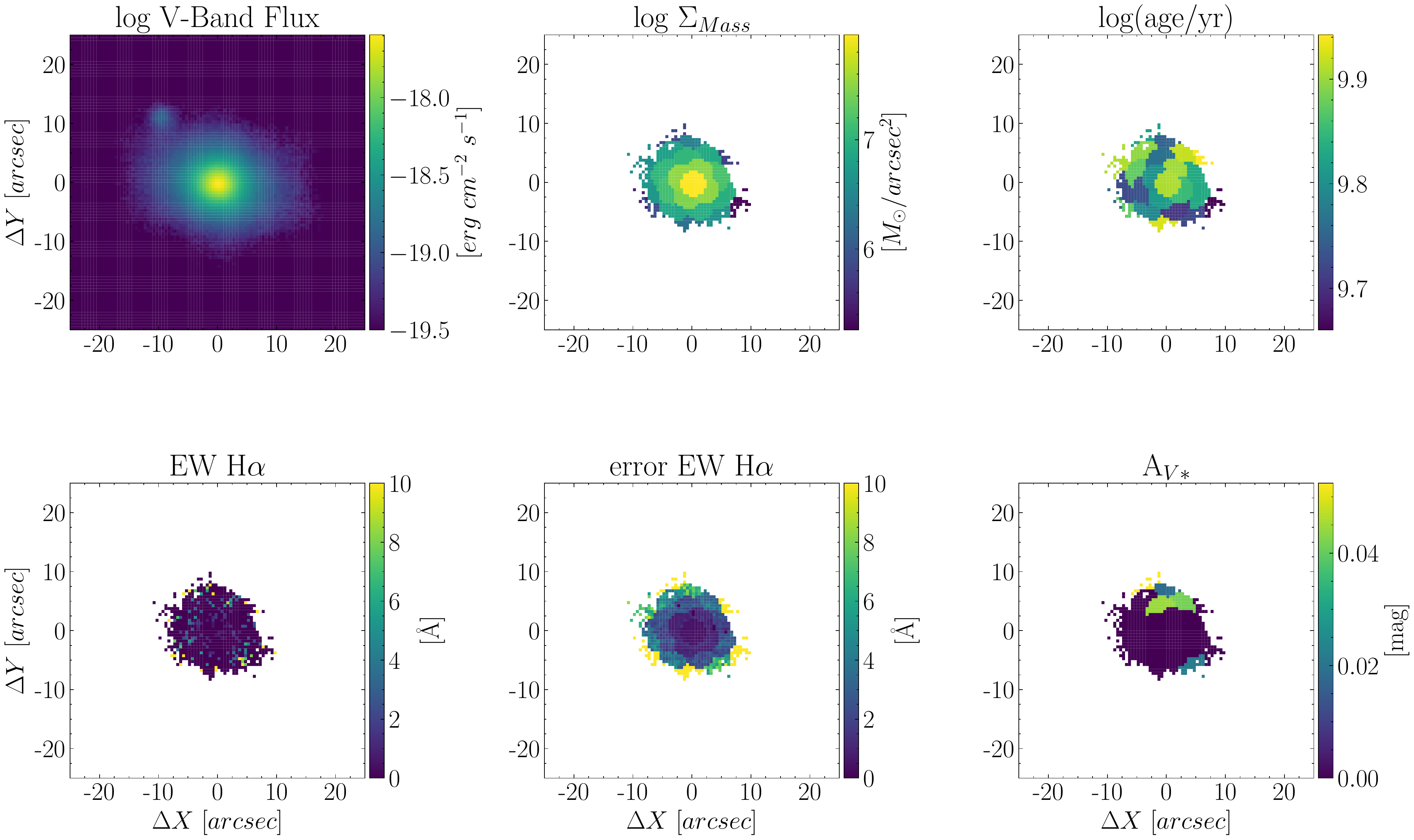}
    \caption{MUSE stellar continuum subtracted emission line maps.}
    \label{fig:muse}
\end{figure}

In Figure \ref{fig:muse} we show the sample of these maps with the main parameters of interest for this study. The galaxy shows a centrally concentrated structure, like most TDE hosts \citep{Hammerstein22}, a very old stellar population (mean age $\geq$ 10$^{9.7}$ yr), a lack of dust ($A_{V*} \leq 0.05$ in all spaxels), and very faint emission lines (mean EW H$\alpha <$  1 \AA), without any apparent increase towards the central spaxels. 

\subsection{X-shooter}
\label{sec:xshooter}
The host galaxy was observed in long-slit mode with the X-shooter instrument on the Very Large Telescope (VLT) Unit Telescope (UT) 3 on 2020 October 16 (MJD 59138.08). Slit widths of 1.0, 0.9 and 0.9 arcsec were deployed for the UVB, VIS and NIR arms, respectively, for a total exposure time of 1300 seconds. 
The average seeing of 0.7 arcsec during the observations results in a seeing-limited spectral resolution of R = 7700 (UVB), 12700 (VIS) and 8000 (NIR), equivalent to a FWHM spectral resolution of 40 km s$^{-1}$ (at 4000 \AA) and 25 km s$^{-1}$ (at H$\alpha$). 
The data were taken in on-slit nodding mode, but to increase the signal to noise ratio (SNR) of the UVB
and VIS arms, we reduce these data using the X-shooter pipeline with recipes designed for stare mode observations. 

\begin{figure}
    \centering
    \includegraphics[width=\textwidth]{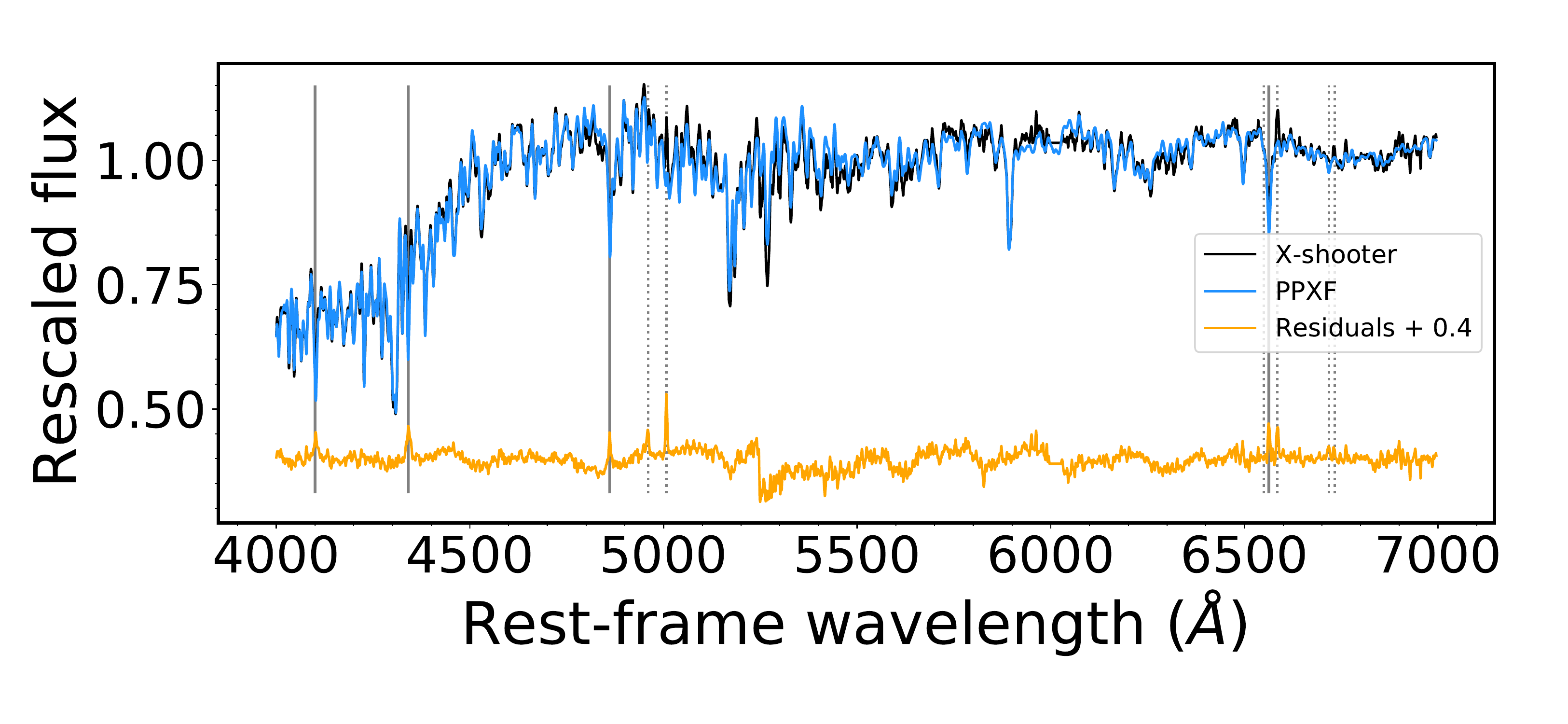}
    \caption{X-shooter spectrum (black), resampled to the native resolution of the MILES stellar template library. The best fit template is shown in blue, and the residuals (offset by +0.4 for clarity) are shown in orange. Weak narrow emission lines appear after template subtraction. Balmer lines are indicated by vertical solid lines, and other emission lines measured for the diagnostic diagrams are marked by dotted lines. The discontinuity between 5300 and 5400 \AA\ is due to the stitching of the two arms.}
    \label{fig:xshooter}
\end{figure}

We modeled the stellar continuum of the X-shooter spectrum with a wavelength range of 4000--7000 \AA\ in the rest-frame, using the penalized pixel fitting (pPXF, \citealt{Cappellari17}) routine. We masked some emission and absorption lines that are usually significant in galaxy spectra since they may affect the best fits of stellar continuum models, e.g., H$\delta$, H$\gamma$, H$\beta$, H$\alpha$, N\,\textsc{ii} 4640, He\,\textsc{ii} 4686, [O\,\textsc{iii}] 4959,5007, He\,\textsc{i} 5875, [O\,\textsc{i}] 6300, [N\,\textsc{ii}] 6548,6584, and [S\,\textsc{ii}] 6717,6731. We used MILES single stellar population (SSP) models \citep{Vazdekis10} as the stellar templates and adopted the SSP model spectra. Given that the initial resolution of the X-shooter spectrum is R$\sim$10000, much higher than that of the MILES spectra (R$\sim$2000), we convolved the X-shooter spectrum to reduce its resolution to R$\sim$2000. Except for the stellar template, a polynomial with degree=4 was added to avoid mismatches between galaxy spectra and stellar templates. The residuals were obtained after subtracting the best-fit stellar continuum model. Residual flux errors are the same as the original flux errors. Line ratios and EWs were measured on the residual spectra, and uncertainties determined by taking into account the flux uncertainties. 
The resampled galaxy spectrum, overlaid with the fit and the residuals after template subtraction are shown in Figure \ref{fig:xshooter}.

%
%

\bibliographystyle{aasjournal}
\bibliography{bib}{}
\end{document}